\documentclass[authoryear]{elsarticle}

\usepackage[english]{babel}
\usepackage{fancyhdr} 
\usepackage{url} 
\usepackage{caption} 
\usepackage{subcaption} 
\usepackage{natbib}
\usepackage{multirow} 
\usepackage{float} 
\usepackage{amssymb}
\usepackage[intlimits]{amsmath} 
\usepackage{amsthm} 
\usepackage[german]{nomencl}
\usepackage{booktabs} 
\usepackage{geometry}
\usepackage{enumitem}

\makenomenclature
\usepackage[pdfborder={000},colorlinks=true,linkcolor=blue,citecolor=blue]{hyperref}
\usepackage{listings}
\usepackage[all]{xy}

\usepackage{color}

\definecolor{b}{rgb}{0,0,.8}	
\definecolor{g}{rgb}{0,.6,0}	
\definecolor{n}{rgb}{0,0,0}	
\definecolor{h}{rgb}{0.4,0.2,0.2}	
\definecolor{v}{rgb}{0.2,0.6,0}

\usepackage[normalem]{ulem}

\DeclareMathOperator*{\argmin}{arg\,min}

\begin{document}

\title{Company classification using machine learning}

\author{Sven~Husmann}
\ead{husmann@europa-uni.de}

\author{Antoniya~Shivarova\corref{cor1}}
\ead{shivarova@europa-uni.de}
\cortext[cor1]{Corresponding author. Tel.: +49 335 5534 2989}

\author{Rick~Steinert}
\ead{steinert@europa-uni.de}
\address{Europa-Universit\"at Viadrina, Gro\ss e Scharrnstra\ss e 59, 15230 Frankfurt (Oder), Germany}


\begin{frontmatter}
\begin{abstract}
The recent advancements in computational power and machine learning algorithms have led to vast improvements in manifold areas of research. Especially in finance, the application of machine learning enables both researchers and practitioners to gain new insights into financial data and well-studied areas such as company classification. In our paper, we demonstrate that unsupervised machine learning algorithms can be used to visualize and classify company data in an economically meaningful and effective way. In particular, we implement the data-driven dimension reduction and visualization tool t-distributed stochastic neighbor embedding (t-SNE) in combination with spectral clustering. The resulting company groups can then be utilized by experts in the field for empirical analysis and optimal decision making. By providing an exemplary out-of-sample study within a portfolio optimization framework, we show that the application of t-SNE and spectral clustering improves the overall portfolio performance. Therefore, we introduce our approach to the financial community as a valuable technique in the context of data analysis and company classification. 
\end{abstract}
\end{frontmatter}

\section{Introduction}\label{sec1:intro}

Machine learning has become one of the most influential phrases in the business world as well as within the recent financial literature. The applications are manifold and include, among others, stock selection \citep{RasekhschaffeJones_2019}, stock prediction \citep{PatelEtAl_2015}, stock trading \citep{DashDash_2016} and portfolio optimization \citep{JainJain_2019}. Moreover, globalization and digital technology allow for the generation and aggregation of financial data in vast amounts, which in turn requires more automated approaches for data analysis. However, the practical implementation of such complicated techniques is often time-consuming and does not coincide with the real challenges an investor has to face with respect to efficiency and performance. Therefore, the correct use of machine learning algorithms necessitates deep knowledge of the underlying mechanics and, more importantly, a functional and straight-forward framework.

In this paper, we combine recent advancements in the field of unsupervised learning with the practical expertise of an investor and introduce a tool for company classification which is easy to implement and works best when used as a decision support module. In particular, we apply the t-SNE algorithm, developed by \citet{MaatenHinton_2008} to decrease the dimension of a high-dimensional financial dataset and thereupon, produce a reasonable and easily-understandable graphical visualization. In a second step, these illustrations can be either visually interpreted by an expert or used within an efficient, data-driven classification analysis to detect groups (or clusters) of similar companies. In a final step, the resulting clusters can be utilized to improve the performance of existing frameworks from the financial research and the practice, for example portfolio optimization. 

In general, we focus on the problem that investors are often confronted with complex and potentially high-dimensional datasets, such as stock returns. To obtain more precise knowledge of such data, it is crucial to reduce the data dimension in a way that increases interpretability and simultaneously minimizes information loss. Up to this date, the principal component analysis (PCA), introduced by \citet{Hotelling_1933}, remains the most commonly used unsupervised learning tool for dimension reduction within the financial literature \citep[see, e.g.][]{PattarinEtAl_2004, RothmanEtAl_2010, FanEtAl_2014, HanGe_2020}. Nonetheless, in other fields, the t-SNE algorithm has recently increased in popularity as a tool for dimension reduction and data illustration \citep[see, e.g.][]{PezzottiEtAl_2017, SchubertGertz_2017, AroraEtAl_2018}. As a result, various applications have been developed, for example in natural science \citep[see, e.g.][]{MacoskoEtAl_2015, LiEtAl_2017, TravenEtAl_2017}. In the financial literature, the application of the t-SNE is yet rare. \citet{KalsyteVerikas_2013} exploit the properties of t-SNE for analyzing the financial soundness of companies. \citet{Sarlin_2015} provides a qualitative overview of data and dimension reduction methods for visual financial performance analysis. \citet{WuEtAl_2019} introduce a deep learning framework for predicting stock prices. Throughout the analysis, they use a two-dimensional t-SNE on the final stock representation to assess and interpret the stocks' universe. Still, to the best of our knowledge, the current financial research exploits t-SNE solely for visualization purposes, whereas our approach includes the dimension reduction feature of the technique directly into existing approaches, e.g. portfolio optimization. 

The low-dimensional data resulting from the application of t-SNE can be either provided to experts for visual and exploratory data analysis, or used within a clustering algorithm for grouping purposes. Especially in the financial literature, company classification plays an important role in the fields of company valuation, financial performance and portfolio optimization \citep[see, e.g.][among others]{KahleWalkling_1996, LinaresMustarosEtAl_2018, FanEtAl_2016a}.  \citet{Farrell_1974} and \citet{Arnott_1980} belong to the first who analyze the covariates and correlations of stock returns to build homogeneous stock groups. In the following decades, however, more sophisticated methods have been investigated in the context of financial time series. \citet{GolosnoyOkhrin_2009} use K-means clustering to group assets and develop a flexible shrinkage estimator for the optimal portfolio weights.  The econophysics literature proposes correlation-based clustering procedures such as hierarchical clustering for filtering the ill-conditioned correlation coefficient matrix \citep[see, e.g.][]{BonannoEtAl_2004, TolaEtAl_2008, TumminelloEtAl_2010, BridaRisso_2010}. \citet{NandaEtAl_2010} compare different clustering algorithms for the application within a portfolio optimization framework. Furthermore, \citet{DePrado_2016} introduces an entirely new method for portfolio optimization, which exploits hierarchical clustering and graph theory. Since it models pairwise relations within data, graph theory has already been successfully applied for the estimation of the covariance matrix of stock returns \citep[see, e.g.][]{FriedmanEtAl_2008, GotoXu_2015}. 

Here, we implement the spectral clustering algorithm by \citet{NgEtAl_2002} as a way to connect the importance of both clustering algorithms and graph theory for identifying groups and relations within stock returns' data. Using a similarity rather than a distance measure, spectral clustering can more efficiently cluster high-dimensional and less compact data. Moreover, since no assumptions are made for the particular shape of the derived clusters, the algorithm can detect non-convex data structures. Both of these properties fit the idea and mathematics behind t-SNE, as shown in \citet{RogovschiEtAl_2017} and \citet{LindermanEtAl_2019}. Additionally, t-SNE can provide the necessary affinity matrix and hence, speed up the otherwise computationally expensive spectral clustering algorithm. Therefore, our approach leads to an effective and efficient classification of company data by combining the dimension reduction properties of t-SNE with spectral clustering. 

To validate the proposed approach, we first show that the clusters resulting from an application of t-SNE on stock returns provide similar results to a proprietary industry classification scheme even if groups are formed just by visual representation. However, if instead of visual representation the data is passed to a clustering algorithm, we can combine the t-SNE plus clustering with any standard model from the financial literature, allowing for machine-based cross-validation (CV) even within problems which naturally do not incorporate any parameter tuning. In this way, our approach can potentially lead to overall performance improvement, especially if the modeler has to estimate certain parameters. We refer to this as a decision engine. We demonstrate within an exemplary portfolio optimization framework that our decision engine yields better results than other standard classification approaches in terms of Sharpe ratio.

The rest of the paper is organized as follows: In Section~\ref{sec2:tSNE}, we introduce the main properties of t-SNE for company classification. Section~\ref{sec3:decision} outlines a decision engine for the optimal selection of algorithm-specific tuning parameters within a general model framework. In Section~\ref{sec4:app}, we provide an exemplary application for portfolio optimization and report the corresponding empirical results. Section~\ref{sec5:concl} summarizes and concludes.

\section{t-SNE for Company Classification}\label{sec2:tSNE}

Due to its property to reduce a dataset's dimension and to provide a good interpretation of underlying data structures, the t-SNE algorithm of \cite{MaatenHinton_2008} is nowadays mostly used as an unsupervised visualization tool for high-dimensional data. In that sense, t-SNE can be compared to the well-known PCA. While PCA achieves dimension reduction by a linear projection, t-SNE utilizes present nonlinear relationships within the dataset. Moreover, PCA aims at retaining only the global structure of the data (the global variance across the entire dataset), whereas t-SNE balances global and local structure. The t-SNE algorithm has, therefore, the benefit that clusters formed in the low-dimensional space are interpretable as data points that were also very similar in the high-dimensional space, making the resulting visualization easier to analyze from a practitioner's perspective. 

The consideration of local and global structure with the t-SNE technique is assured by the following two-step process. First, t-SNE creates a normal (Gaussian) probability distribution over the full dimension of the dataset to measure the similarity between different data point series $x_i$ to $x_j$ by computing the corresponding conditional probability $p_{j|i}$ for each data pair with
\begin{equation}\label{eq_condprob1}
p_{j|i} = \frac{exp\left(-||x_i-x_j||^2/2\sigma_i^2\right)}{\sum_{k\neq i} exp\left(-||x_i-x_k||^2/2\sigma_i^2\right)}\,,
\end{equation}
where $\sigma_i$ refers to the standard deviation of the data series $x_i$, or the so-called bandwidth of the Gaussian kernel. Assuming the neighboring data points are selected in proportion to their probability density under a Gaussian, centered at $x_i$, the similarity of $x_j$ to $x_i$ is the conditional probability $p_{j|i}$ that $x_i$ would pick $x_j$ as a neighbor \citep{MaatenHinton_2008}. As a result, similar data series have a high probability of being picked as a neighbor and vice versa. In a second step, t-SNE maps these probabilities to a lower, usually two- or three-dimensional space by modeling the similarity of map point $y_j$ to map point $y_i$ with the conditional probability $q_{j|i}$ given as
\begin{equation}\label{eq_condprob2}
q_{j|i} = \frac{(1+||y_i-y_j||^{-1})}{\sum_{k\neq i} (1+||y_i-y_k||^{-1})}\,,
\end{equation}
where the new lower-dimensional representations of $x_i$ and $x_j$ are $y_i$ and $y_j$, respectively. The main novelty of t-SNE lies in the definition of the conditional probability $q_{j|i}$. In particular, whereas the original SNE algorithm by \citet{RoweisEtAl_2002} uses a Gaussian distribution  as well in the low-dimensional space, t-SNE employs a Student-t distribution with one degree of freedom, also known as the Cauchy distribution, to solve for the so-called ``crowding problem".\footnote{The crowding problem is defined in \citet{MaatenHinton_2008} as ``the area of the two-dimensional map that is available to accommodate moderately distant datapoints will not be nearly large enough compared with the area available to accommodate nearby datapoints".} Hence, by introducing a heavy-tailed distribution, t-SNE ensures that the data can be spread wider and more distinctly within the low-dimensional (visualization) space. Such a feature is extremely valuable for a dimension reduction tool, which is supposed to work well for high-dimensional data. Furthermore, an exact or perfect representation of the high-dimensional space on the low-dimensional map would imply that the conditional probabilities $p_{j|i}$ and $q_{j|i}$ are equal. Therefore, to achieve the best dimensionality reduction, in a final step t-SNE aims to minimize their difference. A typical measure for the difference between probability distributions is the relative entropy - the Kullback-Leibler (KL) divergence \citep{KullbackLeibler_1951}. Contrary to SNE, t-SNE uses the single KL divergence between the joint probabilities $p_{ij}$ and $q_{ij}$ as a cost function   
\begin{equation}\label{eq_kld}
C=KL(P||Q)=\sum_i\sum_j p_{ij}\log(\frac{p_{ij}}{q_{ij}})\,,
\end{equation}
with the gradient
\begin{equation}\label{eq_klgrad}
\frac{\delta C}{\delta y_i}=4\sum_j\left(p_{ij}-q_{ij}\right)\left(y_i-y_j\right)\,,
\end{equation}
which is much simpler and more computationally efficient than in the case of SNE.\footnote{The cost function within both SNE and t-SNE is minimized with a gradient descent algorithm.} In addition, \citet{Maaten_2014} approximates the gradient by implementing the Barnes-Hut algorithm and proves the outperformance both in results' quality and computational acceleration. Overall, the final step of t-SNE minimizes the divergence between two distributions: a distribution that measures pairwise similarities of the input data and a distribution that measures pairwise similarities of the corresponding low-dimensional points. The resulting two- or three-dimensional data can be easily visualized to represent distinct clusters, referred to as groups, and henceforth, utilized for specific financial applications.

\subsection{Challenges}

Despite its broad applicability and promising properties for dimension reduction and visualization of high-dimensional data, t-SNE does exhibit some drawbacks. One practical problem, originating from the two-step process in Equations~\eqref{eq_condprob1} and \eqref{eq_condprob2}, is that the distances between the resulting data points on the low-dimensional space do not have any meaningful interpretation without further manipulation \citep{WattenbergEtAl_2016}. Contrary to PCA, t-SNE does not provide a functional representation of how to map the data. It is, therefore, necessary to always apply the t-SNE algorithm on the full dataset with all its components. 

Most importantly, the performance of t-SNE depends strongly on the choice of the so-called perplexity parameter, which per definition is an information measure for the prediction inaccuracy of a probability distribution $p_i$ and increases monotonically with the variance $\sigma_i$ of the data series $x_i$. Within t-SNE, the perplexity parameter can be interpreted as a smooth measure for the effective neighbors of $x_i$ and defines how strongly to consider the local and the global aspects of the data. For instance, if the perplexity is chosen to be too low, then the t-SNE algorithm will incorporate more of the local data structure. This initial setting is useful for more dense data with lower variance. Otherwise, it would result in too many groups with a lower inter-group dissimilarity. In contrast, if the perplexity is chosen too high, relevant groups cannot be detected, as the algorithm puts too much weight on the global data structure. The corresponding groups would display a lower intra-group similarity. On the contrary, an adequate dimension reduction technique should result in both high intra-group similarity and inter-group dissimilarity. 

\begin{figure}[h!]
	\centering
	\includegraphics[scale=0.35]{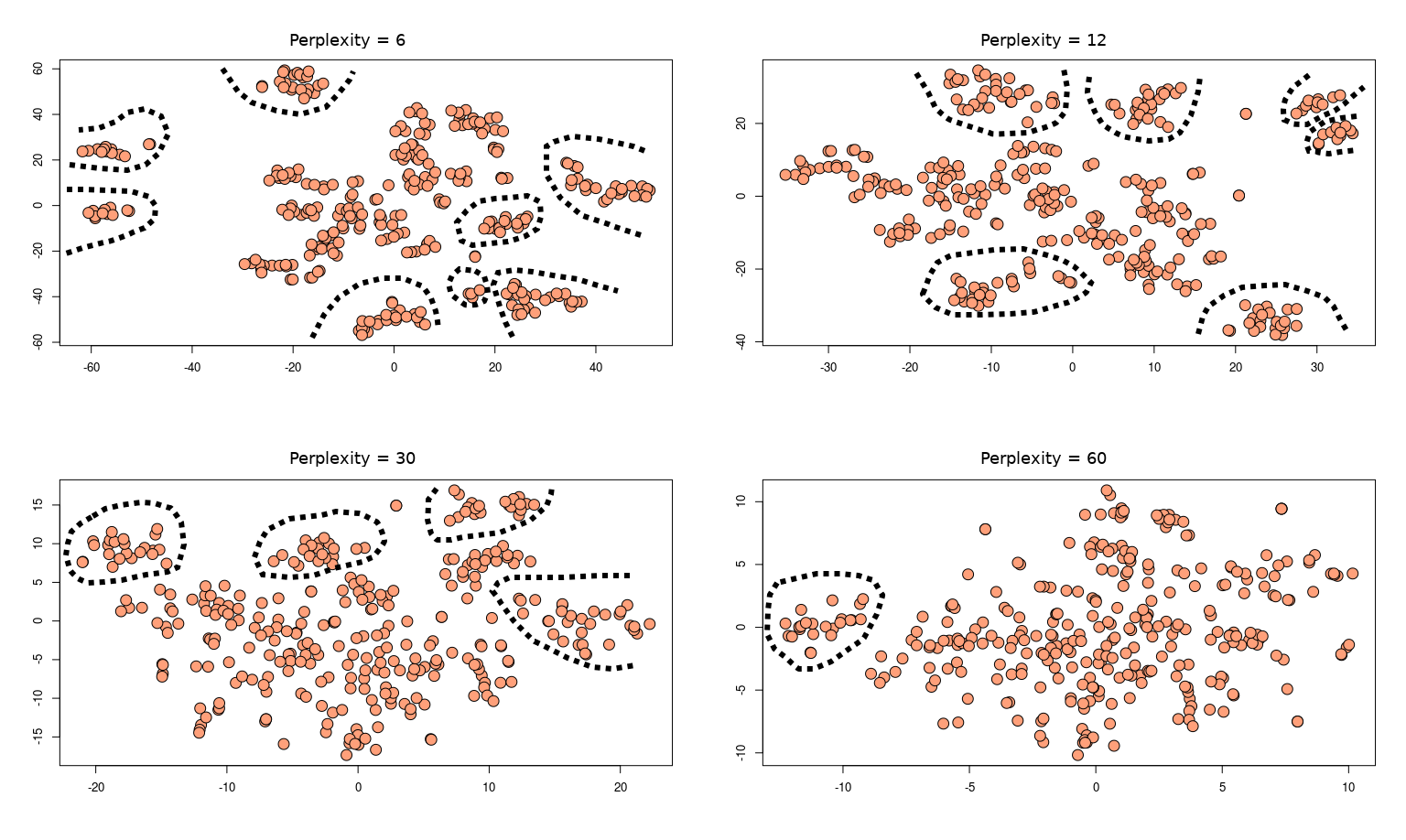}
	\caption{Visualization of companies after applying a two-dimensional t-SNE with different perplexities.}
	\label{fig:tsne_perp}
\end{figure}

To demonstrate the outlined challenges, we exemplary show different outcomes of the t-SNE application on daily returns of 318 S\&P500 companies in Figure~\ref{fig:tsne_perp}. For creating the pictures from top left to bottom right, we use the stock returns of 318 companies from the S\&P500 index over a four-year period from 01/14/14 to 12/31/18 and apply t-SNE for dimension reduction on this dataset. Each orange point in the pictures corresponds to the transformed values $y_i$ and therefore represents one company. The main idea is, that based on the proximity of certain points, arranged to clusters, the user can now refer to companies that are close together as a group. This can be done either manually, by visual interpretation, or automatically, in a data-driven way. For the purpose of illustration at this point, within each picture, we perform a manual grouping by circling data points or companies, which visibly form a distinct cluster. For each picture, we use the same dataset but yet various perplexities within t-SNE, namely 6, 12, 30, and 60, respectively. It is obvious, that the different choice for perplexity leads to strongly different outcomes. In the case of a high perplexity, e.g. 60, only one distinct group can be found. However, for low perplexities such as 6, one can find eight groups and more. As argued beforehand, it can be easily observed from Figure~\ref{fig:tsne_perp} that the perplexity parameter influences the amount of visibly detectable groups. For lower perplexity levels, there are more visible groups, since the algorithm incorporates more of the local data structure, and vice versa.

Since hyperparameters influence strongly the results of t-SNE, to properly exploit its benefits, practitioners and researchers should use it as an effective decision support module - either by additionally incorporating expert knowledge of the data, for example, which groups do indeed make sense, or by applying a proper data-driven validation method such as cross-validation. 

\subsection{Classification}

In a real-life scenario, after applying t-SNE on a high-dimensional data with different perplexities, an investor or a researcher could determine which parameter to use by examining the groups of each outcome. That could be achieved either in accordance with individual practical experience or industry knowledge or by applying a data-driven classification (clustering) algorithm. In case no knowledge of the underlying groups is present or assumed, clustering algorithms can be applied for a data-driven classification. Technically, not all clustering algorithms can utilize t-SNE. In general, the output of t-SNE does not keep the information of the original data density which can be challenging, especially for the class of density-based clustering methods such as DBSCAN \citep{EsterEtAl_1996}. Moreover, since t-SNE allows for various types of cluster shapes, methods such as the K-means clustering \citep{MacQueenothers_1967} may not suffice to build groups and categorize the output in a meaningful way.\footnote{In detail, K-means aims to identify groups by calculating measures such as the Euclidean distance between different points towards a so-called ``centroid". This leads to promising results only when the data is clustered mostly spherical.} In the case of t-SNE, however, as shown in Figure~\ref{fig:tsne_perp} and desired in the case of high-dimensional financial data, the clusters can have a multitude of shapes. This property of t-SNE makes it an effective choice for dimensionality reduction of heterogeneous high-dimensional data such as stock returns but limits its applicability in combination with centroid-based clustering algorithms.

Up to this point, the literature on t-SNE considers graph-based clustering algorithms, in particular, the spectral clustering algorithm \citep[see, e.g.][]{XieEtAl_2016, RogovschiEtAl_2017, LindermanSteinerberger_2019}. The main idea here is that, before the actual clustering process,  spectral clustering constructs an undirected graph, represented by its so-called adjacency matrix $A$, where $A_{ij}$ measures the similarity (or affinity) between the data pair $x_i$ and $x_j$. Only thereafter, the algorithm calculates the corresponding Laplacian matrix $L=D-A$, where $D=\sum_j A{ij}\,$, and utilizes its $k$ largest eigenvalues to carry out the clustering analysis in the feature space constructed by the respective eigenvectors.\footnote{For more details on the application of Laplacian eigenmaps and matrices, see, e.g. \citet{BelkinNiyogi_2003}.} Although spectral clustering is suitable for a dataset with arbitrary shape and high-dimension, the construction of the affinity matrix can be computationally expensive. This is where t-SNE can be applied as a dimension reduction tool. This combination of the two algorithms is supported additionally by \citet{LindermanSteinerberger_2019} who perform a rigorous analysis on the t-SNE mathematical foundations and prove the connection to Laplacian eigenmaps and matrices. Following this line of research, we use the spectral clustering algorithm as implemented by \citet{KaratzoglouEtAl_2004}. 

\begin{figure}[h!]
	\centering
	\includegraphics[scale=0.5]{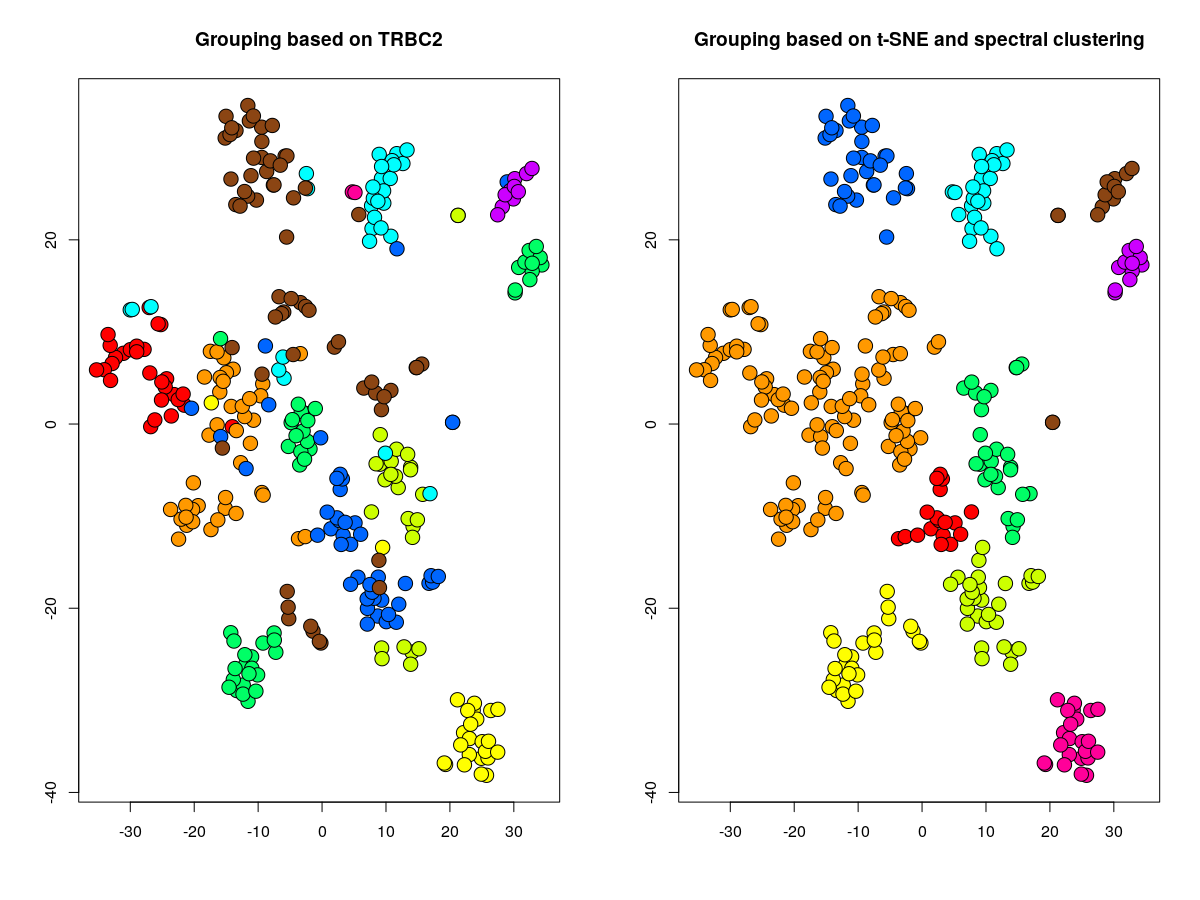}
	\caption{Visualization of companies based on a two-dimensional t-SNE with a perplexity of 12. The colors of the points correspond to a grouping, performed with the Thomson Reuters 2-digit industrial code (left) and a classification using spectral clustering with 10 groups (right).}
	\label{fig:tsne_groups}
\end{figure}

To provide some insight into the quality of the performed grouping from combining t-SNE with the spectral clustering algorithm, in Figure~\ref{fig:tsne_groups} we show the established groups with an exemplary perplexity parameter of 12. On the right-hand side of Figure~\ref{fig:tsne_groups} every point is colorized according to the result of the spectral clustering algorithm, applied on the output of the t-SNE. On the left-hand side, we visualize the results of a proprietary industry classification system as a benchmark. 

By using the first two digits of the Thomson Reuters Business Classification (TRBC2) for each company, we depict the companies from the same industry according to TRBC2 with the same color - for example, every teal-colored point represents a company which belongs to the same group, independent of its position in the two-dimensional graph. Here, we can notice that the grouping, based on visual analysis of the proximity between data points, would indeed lead to groups, which seem to cluster companies within the same industry. However, we point out that the Thomson Reuters Codes are not what is referred to as ``ground truth" - the only real and true grouping for the considered companies. In fact, the real grouping is always unknown and constantly changing due to changes in companies' policies, market size, and other factors. For the application of the spectral clustering algorithm, we set the group size to 10, so that the results can be compared fairly to the ones, produced by the industry classification with TRBC2. We can clearly observe that spectral clustering performs the grouping more consistently than the industry classification and more similar to an investor who visually differentiates between the mapped data points. Still, an advantage of the spectral clustering algorithm is that it performs the company grouping data-driven as opposed to a possibly subjective assessment by an investor.

\subsection{Validation}

After applying either a visual or data-based classification, the resulting groups might be checked for plausibility by an expert. As the ground truth of the underlying groups is unknown, expert knowledge about the business models and/or companies' sectors can be applied to validate or change the group membership of companies. In the case of the dataset at hand, one could, for instance, argue why the group, colored in green (identified with the code 55, referring to ''Financials``) in the left-hand side of Figure~\ref{fig:tsne_groups}, is spread around three distinctive clusters, while in the right-hand side of the pictures, these seem to have different groupings, colored in purple, orange, and yellow, respectively. Thus, the application of t-SNE indicates that the TRBC Financials group forms sub-groups based on their multidimensional similarity profile. The spectral clustering acknowledges this by giving them three distinct colors, for example, one subset forms the purple group in the upper right corner. After manually studying the business profile of the companies in question, we discover that the purple dots all belong to the real estate industry, while the others are mostly banks and investment companies. This shows that the combination of t-SNE and spectral clustering, as opposed to the industry codes TRBC2, could indeed provide further insight into the grouping. However, the yellow group produced by the t-SNE with spectral clustering seems to include companies that are not found in the 55 industry code of TRBC2. Analyzing them shows that this divergence is likely due to a wrong classification, as these are all automotive-related companies. An expert would have easily identified these as wrongfully grouped and thus discarded them. Without the presence of such an expert, the investor can still use data-driven approaches to adjust the grouping by tuning different parameters - for instance, by setting different starting points for the algorithm or changing the number of classes for the spectral clustering.

\begin{figure}[h!]
\centering
\includegraphics[scale=0.5]{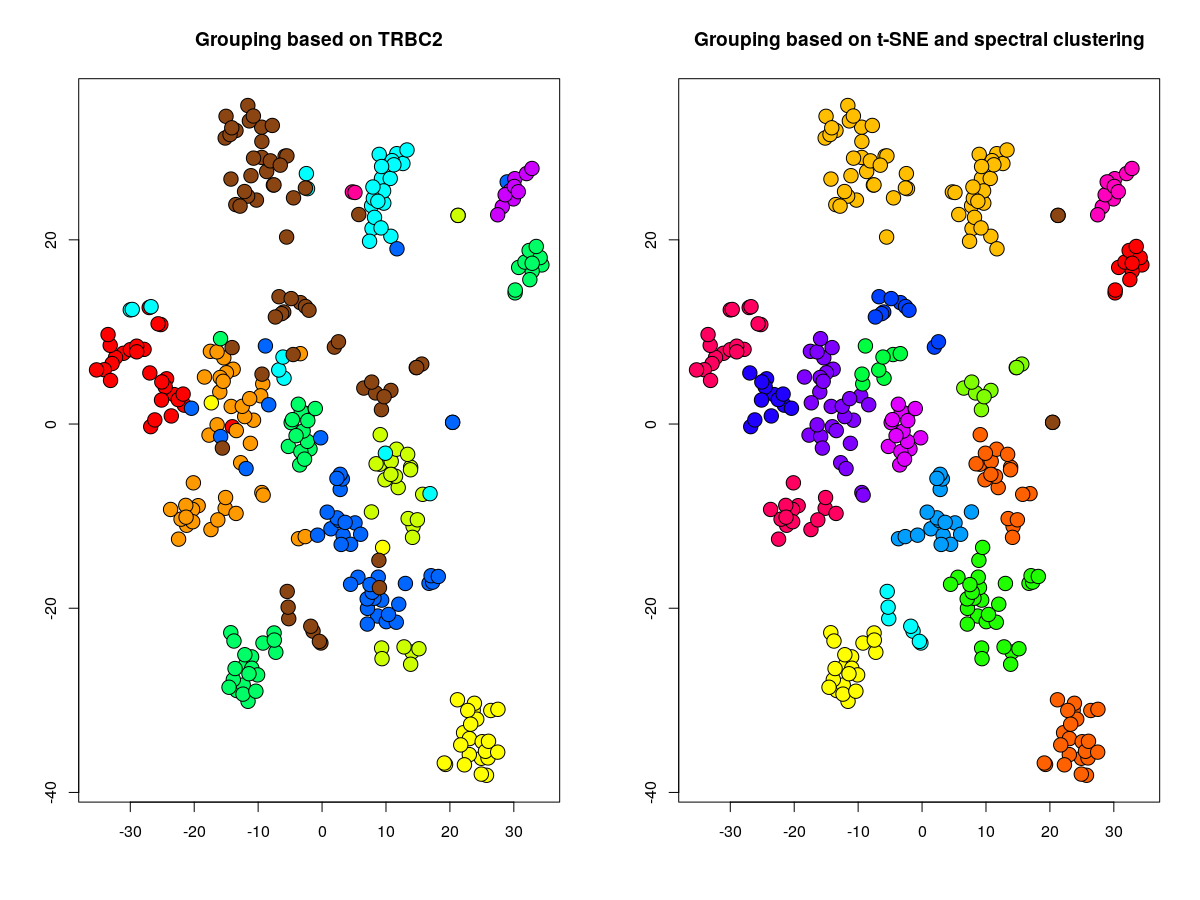}
\caption{Visualization of companies based on a two-dimensional t-SNE with perplexity of 12. The colors of the points corresponds to a grouping done by the Thomson Reuters 2-digit industrial code (left) and a classification using spectral clustering with 16 groups (right).}
\label{fig:tsne_groups_adj}
\end{figure}

We perform exactly these adjustments, i.e. new starting points and a group size of 16, and display the results in Figure~\ref{fig:tsne_groups_adj}. It can now be recognized that the grouping for the financial companies is more accurate and in accordance with the visual representation of the t-SNE.

Still, even if some tuning parameter setting for t-SNE with spectral clustering might match the results from classification with the TRBC-Codes, such grouping is not necessarily beneficial for the usage within, for instance, a portfolio optimization framework. Moreover, a group formation merely on visual analysis by an expert and the general practical expertise may suffer from confirmation bias and subjective reasoning. Accordingly, here we argue that t-SNE becomes especially relevant, if the ground truth of the underlying grouping of some or all the companies is unknown. Examples include groupings by industry, sustainability, profitability. or any similar artificial ranking. In that case, t-SNE can help identify the relevant cluster of that company by simply using similarities in the provided dataset, e.g. return data, balance-sheet data, social responsibility measures, and others. Additionally, missing the appropriate grouping data is often a practical issue, as proprietary data such as the TRBC is often not available for smaller, not traded companies. In contrast, other information, like the balance sheet, is easily accessible. Furthermore, the application of a fully data-dependent approach such as t-SNE with spectral clustering allows for even more precise tuning and quantitative improvement of the performed grouping. To achieve that, we show how to connect the t-SNE plus spectral clustering with any standard approach in a general decision engine.

\section{Decision Engine for Optimal Grouping}\label{sec3:decision}

Besides adjusting the parameters of the proposed algorithms to better match the visible clusters, the approach can be improved in a data-driven way with respect to a certain goodness-of-fit. The choice of the goodness-of-fit criterion depends on the original purpose of the study or practical application and can be, for instance, the mean-squared-error (MSE), the penalized maximum-likelihood or the minimum-variance (MV). In addition, the perplexity parameter within t-SNE, as well as the group size within spectral clustering can both be regarded as tuning parameters, which, when changed, lead to deviating results. Moreover, due to the randomization of starting points for both the t-SNE as well as the spectral clustering algorithms, outcomes differ in general. Such parameter dependency and possible randomness of the results provide the opportunity and somewhat necessity, to use cross-validation to detect the optimal parameters and hence, the optimal grouping. 

The main advantage of such a data-driven approach is its flexibility in terms of a model setup. For example, in the practically relevant case of company valuation with multiples \citep[see, e.g.][]{Schreiner_2009} the investor is obliged to define peer-groups based on similar idiosyncratic risks. In this case, groups should be detected so that there is high homogeneity within each group. While choosing too many groups leads to high correlations and thus a lack of dissimilarity between them, choosing too few groups leads to groups that contain companies with too diverse and incomparable business models. Another example from finance is portfolio optimization. From a theoretical standpoint, a portfolio of stocks is to be optimized with the least amount of groups as possible (preferably only one group), so that the desired diversification effect can be achieved \citep[see, e.g.][]{Markowitz_1952, GreenHollifield_1992, DomianEtAl_2007}. In practice, on the other hand, this scenario yields large concentration ratios with comparably many companies per observation point. The negative effect of the curse of dimensionality can then negate the positive diversification effect due to high estimation error in the necessary parameters - the expected returns and the covariance matrix of returns \citep{Michaud_1989, ChopraZiemba_1993, JagannathanMa_2003}. Considering the trade-off between estimation risk and diversification, the optimal group size can, therefore, be established by a data-driven grouping optimization. 

\begin{figure}[h!]
\centering
\includegraphics[width=1\linewidth, height=6cm]{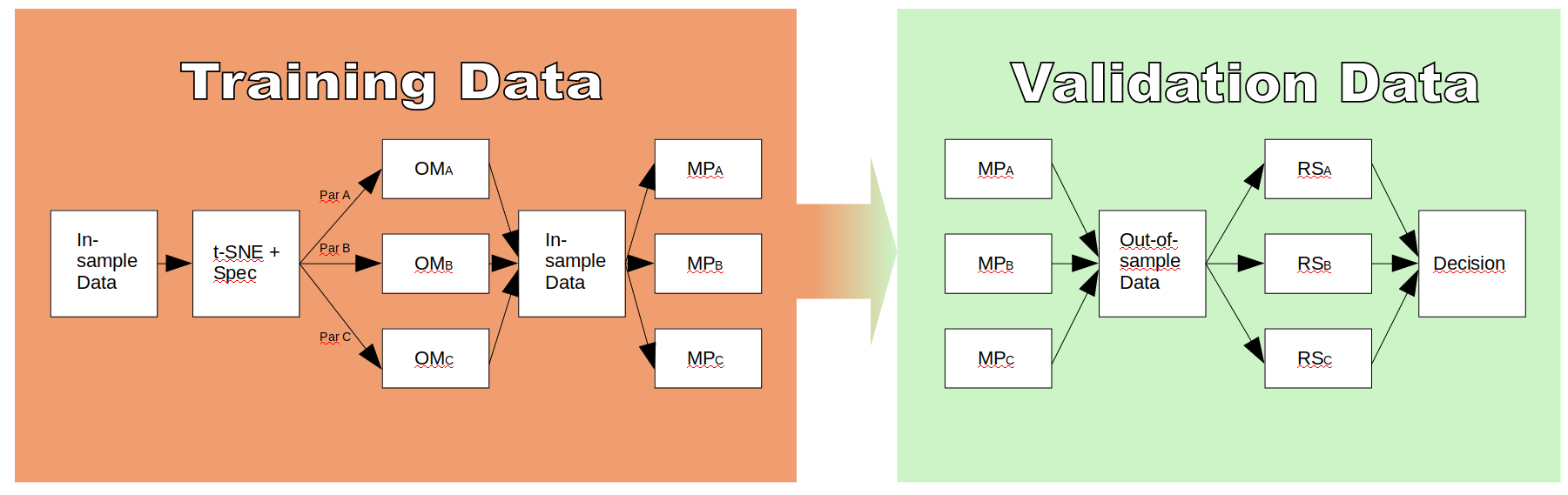}
\caption{Decision engine for the incorporation of t-SNE and spectral clustering (spec) into another modeling approach. OM stands for original model, MP for model parameters, RS for results.}
\label{fig:dec_eng}
\end{figure}

Since the purpose of an empirical investigation is usually aiming to optimize a parameter of interest, for instance MV for minimum-variance portfolios or MSE for company valuation, our data-driven machine learning tool can add value to a general optimization framework by introducing a new layer of decision and optimizing with respect to the grouping quality. Aiming at designing an easy-to-implement framework for company classification with machine learning, in Figure~\ref{fig:dec_eng} we present the decision engine for our approach. 

The decision engine requires the data to be split into a training dataset, which is also referred to as in-sample data, a validation dataset, which is also referred to as out-of-sample data, and a test dataset, used for the final evaluation.\footnote{See, e.g. \citet{Hjort_1996} for more information on the terminology.} As depicted on Figure~\ref{fig:dec_eng}, from left to right, we firstly use the in-sample data and pass it to the t-SNE with spectral clustering approach. Based on a predefined parameter set, for instance, for perplexity, group size, etc., there is a variety of different results for the performed grouping. For illustration purposes, in Figure~\ref{fig:dec_eng} we show overall three different scenarios with parameter sets Par A, Par B and Par C, respectively. The generated grouped data can then be used within the original model (OM). If the model, as in the case of company valuation, requires already grouped companies, the resulting groups can be directly applied. However, if the original model, as often the case, does not incorporate grouping, a slight adjustment is necessary. One of the simplest ways for adaptation is to apply the OM to each of the groups for a certain parameter and later aggregate the results with a suitable measure, for example, the average. Applying the OM yields the necessary model parameters (MP). However, for practically relevant performance evaluation, it is necessary to figure out which model performs best on a new dataset. This is shown on the right-hand side of Figure~\ref{fig:dec_eng}, where the MPs are now directed to the out-of-sample dataset, keeping the grouping step of earlier intact. The respective results (RS) can be drawn from any measure-of-fit and give information about each model's performance, conditional on the chosen parameters. The parameter set which yields the best results is then chosen as the best model, providing the investor overall not only a potentially better fit to the data but also a suitable choice of grouping, given the problem at hand. Using the proposed decision engine multiple times, by splitting the dataset repeatedly in different training and validation datasets and averaging over all results or decisions, results in the standard CV approach for parameter choice. The final chosen model is then applied to the test dataset, which is used neither in the training nor in the validation of the model. This test dataset, if chosen appropriately, is therefore used to reveal the true predictive power of the chosen model.

The advantage of this approach is that original models, which usually do not have tuning parameters, such as the group size, can now be fine-tuned to provide a better fit to the data. Generally speaking, applying an unsupervised machine learning algorithm for grouping enriches the model variety and thus provides an opportunity for more extensive model evaluation. To show the wide range of applicability and the performance of our framework for company classification with t-SNE, spectral clustering, and CV, we present an exemplary application from the field of portfolio optimization. 

\section{Application for Portfolio Optimization}\label{sec4:app}

One of the most important challenges in finance is the problem of optimizing a portfolio. However, due to uncertainties in the estimation of input parameters, the expected returns and the covariance of asset returns, the efficient estimation of optimal portfolio weights for each asset remains a challenge. Since Markowitz postulated what is nowadays known as Modern Portfolio Theory \citep{Markowitz_1952}, financial researchers apply methods from various scientific fields for implementing portfolio strategies in practical scenarios. Some penalize the optimal weights by using practically relevant constraints or some form of regularization \citep[see, e.g.][]{JagannathanMa_2003, BrodieEtAl_2009, DeMiguelEtAl_2009, FanEtAl_2012a, HusmannEtAl_2019}. Others focus on directly improving the estimation of the expected returns \citep[see, e.g.][]{Jorion_1991, BestGrauer_1991} and the covariance matrix of returns \citep[see, e.g.][]{ LedoitWolf_2004, LedoitWolf_2017b, FanEtAl_2013}. \citet{DeMiguelEtAl_2009a} even show that the estimation-free, equally-weighted portfolio can be superior to the Markowitz portfolio in terms of return-risk ratio, the so-called Sharpe ratio. Another line of research considers stock grouping as the appropriate strategy to reduce the underlying data dimension and therefore, the estimation error due to the curse of dimensionality \citep[see, e.g.][]{NandaEtAl_2010, GuptaEtAl_2011, LongEtAl_2014, DePrado_2016, HanGe_2020}. Our approach extends the financial research about optimal grouping for portfolio optimization by utilizing state-of-the-art unsupervised machine learning techniques.

In detail, we apply t-SNE with spectral clustering for company classification to improve the performance of the tangency portfolio. We use the data-driven approach as described in Section~\ref{sec3:decision} and the out-of-sample Sharpe ratio as a goodness-of-fit criterion. As a next step, we proceed with the CV-based optimization, which has been recently proven to be especially useful for minimum-variance portfolios \citep{DeMiguelEtAl_2013, BanEtAl_2018, HusmannEtAl_2020}. Nonetheless, in the standard Markowitz portfolio optimization framework there is no need for a CV, as it is assumed that either all necessary parameters are known or at least can be sufficiently estimated. Hence, the application of our decision engine can yield additional improvement of the results by selecting the tuning parameters in a sophisticated manner. 

\subsection{Empirical Set-Up}

To incorporate effectively our approach and hence, a company classification into the standard portfolio optimization problem, we need to slightly adjust the general optimization problem, defined as
\begin{eqnarray}\label{eq_tangport}
\widehat{w} = \argmin_{w}
\label{eq_general_frame_full1}  
&& 
w' \widehat{\Sigma} w 
\label{eq_general_frame_objective} \\	
\text{s.t.}
&&
1' w = 1  
\label{eq_general_frame_st_1a} \\
\text{}
&&
\mu' w = \mu_f,
\end{eqnarray}
where $\widehat{w}$ is the estimated weights vector for the assets, $\widehat{\Sigma}$ is the estimated covariance of returns, $\mu_f$ is a target expected return, and \eqref{eq_general_frame_st_1a} is the sum constraint, which ensures that all weights sum up to 1. Although this optimization is left untouched, following Section~\ref{sec3:decision}, we can easily pre-process the companies' return data with t-SNE to gain relevant, low-dimensional grouped data. Recognizing that the weight parameter $\widehat{w}$ can be applied to any asset, such as single stocks but as well as portfolios, we can adjust the underlying dataset in a meaningful way. The standard case, where any asset is treated individually, can be considered the special case of a grouping with the number of groups equal to the number of assets. Thus, if we group all companies into smaller subgroups we can technically treat each group as one asset, i.e. a (sub-)portfolio, as well. In order to keep the analysis simple and to avoid unnecessary estimation error, we create such a group-based sub-portfolio by assigning to each company the weight $\frac{1}{N_g}$, where $N_g$ is the number of companies in that group. Hence, we construct an equally-weighted (also called naive) portfolio on the intra-group level. On these resulting portfolios, ranging from 1 (no grouping at all) to $p$ (each company is its own group with $p$ companies in the dataset), we then construct the tangency portfolio as in Equation~\eqref{eq_tangport}. Since our aim is to emphasize the effect of grouping in portfolio optimization, we use this procedure for a predefined set of groups, ranging from $g=2$ to $g=20$. 

\begin{figure}[h!]
\begin{subfigure}{.5\textwidth}
  \centering
  \includegraphics[width=.95\linewidth]{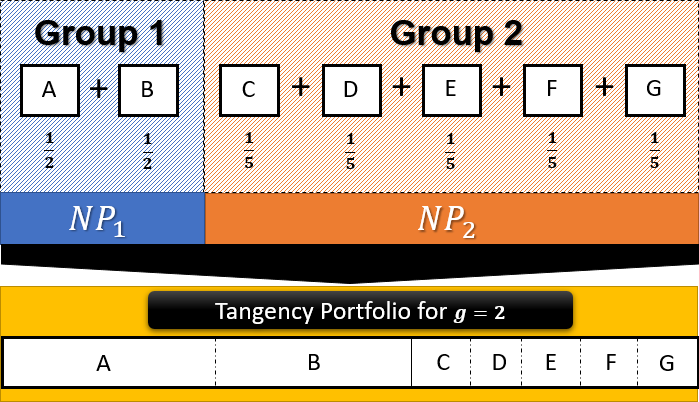}
  \caption{Portfolio formation with $g=2$}
\end{subfigure}%
\begin{subfigure}{.5\textwidth}
  \centering
  \includegraphics[width=.95\linewidth]{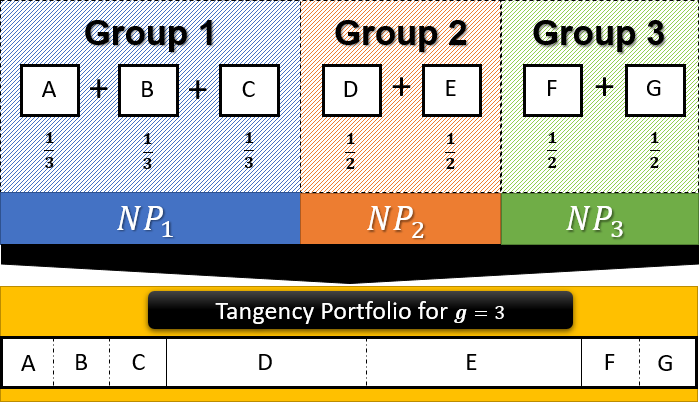}
  \caption{Portfolio formation with $g=3$}
\end{subfigure}
\caption{Using the t-SNE with spectral clustering to create naive portfolios out of grouped companies which subsequently will be transformed into one global tangency portfolio. The letters A to G represent different companies, which are first combined into a naive portfolio $NP_i$ for each of the corresponding groups $i$. These portfolios are then combined to one tangency portfolio, which keeps the weights within the group equally sized, but adjusts the weights of each naive portfolio in the final portfolio decision. This is represented by the lower bar in each of the figures.}
\label{fig:portfolios}
\end{figure}

Figure~\ref{fig:portfolios} depicts the described process of first creating naive portfolios, given the grouping decision of the t-SNE and spectral clustering approach, and afterwards forming a tangency portfolio out of all naive portfolios. 
Following the set-up, the number of naive portfolios always coincides with the number of groups. Both sides of Figure~\ref{fig:portfolios} illustrate the difference of the result for seven companies when $g=2$ for the left-hand side and $g=3$ for the right-hand side. Since the group size can constantly change due to the data-driven character of the decision engine, the size and constitution of the naive portfolios can be different for each fixed amount of groups $g$. We can also observe that the algorithm sometimes tends to choose a large cluster of companies as one group and leaving only a few other companies together in another group. However, as all naive portfolios will be passed to the standard optimization, described in Equation~\eqref{eq_tangport}, each asset will receive its final weight based on the membership to a certain naive portfolio and thus depending on the number of groups $g$. As outlined in Section~\ref{sec3:decision}, for each $g$ we create groups depending on a predefined parameter set. In our study, we use a grid of parameters based on the perplexity parameter as well as on the randomness of the spectral clustering. For each perplexity parameter in $S=\{3,8,13,\ldots,53\}$, we simply apply the spectral clustering 30 times, leading to overall $11*30=330$ different grouping outcomes per predefined group size $g$. This approach is used to address the issue of randomness in the spectral clustering. All of these 330 models will then be passed through our decision engine by combining the different companies into an equally-weighted portfolio for each group and finally constructing the tangency portfolio out of the resulting portfolios. The size of the training dataset is $5*252=1260$ observations and the validation set consists of $252$ observations. We use the returns of the validation set to decide on the final model, based on the out-of-sample Sharpe ratio, evaluated as $\frac{\mu}{\sigma}$, whereas $\mu$ is estimated by the sample mean and $\sigma$ by the sample standard deviation. 


To create a large test dataset, we incorporate our decision engine in a standard daily rolling window study. We set our initial test data point to the day which directly follows the validation data and report the returns of our chosen portfolio, depending on the group size $g$. We then shift the rolling window by one day, adding the former test data point to the full dataset, deleting the first observation, and calculating again the return on the new test data point. Every 252 trading days we reapply our decision engine to adjust our classification and adapt to new market conditions. Overall, we shift our rolling window exactly 1259 times, leading to 1259 daily test data points, for which we can calculate the true daily Sharpe ratio an investor would have achieved by applying our framework. 

In total, we use 318 companies listed in the S\&P 500 from 12/27/11 to 12/31/18, downloaded from the EIKON database of Thomson Reuters. We use only those companies, for which the price data is available for the whole period. We use discrete returns and assume that the risk-free rate of return $r_f$ is zero for every period. The number of observations available for estimating the portfolio weights as in Equation~\eqref{eq_tangport}, is approximately two trading years, i.e. $2*252=504$. This setting ensures the realization of a portfolio optimization even in the case of 318 groups (each company is a group on its own).

\subsection{Empirical results}

To examine thoroughly the performance of our approach, from now on referred to as $TS_g$, we introduce some strong, relevant benchmarks. First, the standard tangency portfolio as the global market portfolio $MW_{full}$. It is constructed by simply applying Equation~\ref{eq_tangport} to all companies. Furthermore, as argued by \citet{DeMiguelEtAl_2009a}, a naive portfolio of the same companies does not suffer from estimation error and hence, can serve as a powerful benchmark as well. We refer to the naive portfolio on all stocks as $N_{full}$. However, these two approaches do not consider the grouping of the data and thus, could be considered unfair benchmarks. Therefore, we introduce three more benchmarks, based on different grouping strategies. First, we use the Thomson Reuters industry classification system with the two leading digits of the code, form groups accordingly and apply the decision engine as in Figures~\ref{fig:dec_eng} and~\ref{fig:portfolios} to each group. This is referred to as $TR2$. Secondly, we again use the Thomson Reuters codes but this time with four digits. We call this benchmark $TR4$. Finally, we construct a random grouping strategy, based on the group sizes of the t-SNE and spectral clustering approach for all different predefined group sizes and again apply the methods of Figures~\ref{fig:dec_eng} and~\ref{fig:portfolios}. By randomly allocating companies to specific groups, we can determine the effect of the t-SNE in combination with spectral clustering, when all other relevant factors are held constant. To stabilize the results of the random grouping, we repeat it exactly 100 times with different company allocations and calculate the average Sharpe ratio for each number of groups. This method is referred to as $RND_g$. 
\begin{figure}[h!]
\centering
\includegraphics[scale=0.8]{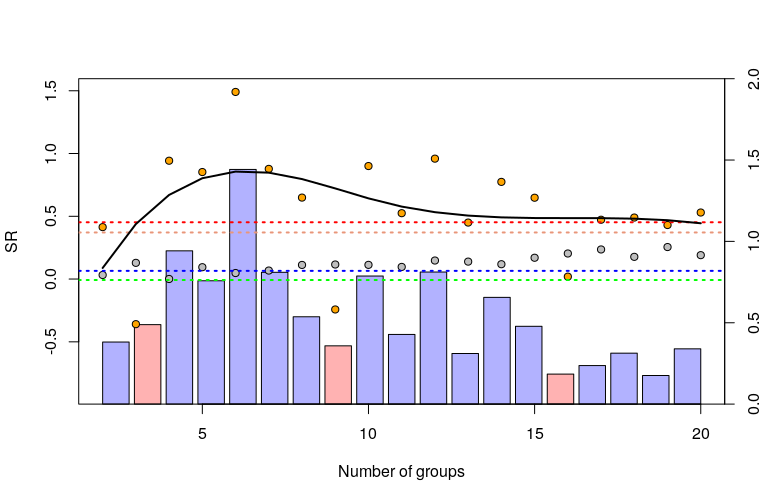}
\caption{Annualized Sharpe ratio (left ordinate) of the test data returns, sorted by the amount of groups $g$. Orange dots represent the corresponding Sharpe ratio of the t-SNE and spectral clustering approach $TS_g$,´ gray dots represent the Sharpe ratio of the grouping approach using random grouping $RND_g$. Bars and bar colors indicate whether the proposed method $TS_g$ yields better (blue) results or worse (red) than $RND_g$. The absolute difference of both models is depicted at the right ordinate. The dashed lines represent the benchmarks for $MW_{full}$ (red), $N_{full}$ (salmon), $TR2$ (blue) and $TR4$ (green). The black line is a trend line for the results of the t-SNE with spectral clustering approach.}
\label{fig:res}
\end{figure}
\begin{table}[h!]
	\resizebox{\textwidth}{!}{
		\begin{tabular}{  ccccccc|cccccc} 
			$g$ & $TS_g$ & $RND_g$ & $MW_{full}$ & $N_{full}$ & $TR2$ &$TR4$& $TS_g$ & $RND_g$ & $MW_{full}$ & $N_{full}$ & $TR2$ &$TR4$\\ \hline
			&\multicolumn{6}{c|}{Sharpe ratio}&\multicolumn{6}{c}{Standard error} \\ \hline
			1 & - & - & 0.452 & 0.371 & - & - & - & - & 0.270 & 0.441 & - & - \\ 
			2 & 0.414 & 0.032 & - & - & - & - & 0.403 & 0.424 & - & - & - & - \\ 
			3 & -0.359 & 0.129 & - & - & - & - & 0.440 & 0.439 & - & - & - & - \\ 
			4 & 0.943 & 0.001 & - & - & - & - & 0.376 & 0.428 & - & - & - & - \\ 
			5 & 0.853 & 0.094 & - & - & - & - & 0.336 & 0.419 & - & - & - & - \\ 
			6 & 1.491 & 0.048 & - & - & - & - & 0.462 & 0.405 & - & - & - & - \\ 
			7 & 0.878 & 0.068 & - & - & - & - & 0.438 & 0.438 & - & - & - & - \\ 
			8 & 0.649 & 0.112 & - & - & - & - & 0.326 & 0.380 & - & - & - & - \\ 
			9 & -0.242 & 0.116 & - & - & - & - & 0.606 & 0.368 & - & - & - & - \\ 
			10 & 0.900 & 0.113 & - & - & -0.008 & - & 0.406 & 0.386 & - & - & 0.484 & - \\ 
			11 & 0.525 & 0.097 & - & - & - & - & 0.326 & 0.420 & - & - & - & - \\ 
			12 & 0.959 & 0.148 & - & - & - & - & 0.460 & 0.394 & - & - & - & - \\ 
			13 & 0.450 & 0.139 & - & - & - & - & 0.459 & 0.423 & - & - & - & - \\ 
			14 & 0.774 & 0.118 & - & - & - & - & 0.451 & 0.424 & - & - & - & - \\ 
			15 & 0.648 & 0.170 & - & - & - & - & 0.360 & 0.419 & - & - & - & - \\ 
			16 & 0.019 & 0.204 & - & - & - & - & 0.555 & 0.450 & - & - & - & - \\ 
			17 & 0.473 & 0.236 & - & - & - & - & 0.325 & 0.422 & - & - & - & - \\ 
			18 & 0.490 & 0.177 & - & - & - & - & 0.443 & 0.389 & - & - & - & - \\ 
			19 & 0.431 & 0.255 & - & - & - & - & 0.385 & 0.376 & - & - & - & - \\ 
			20 & 0.530 & 0.190 & - & - & - & - & 0.502 & 0.446 & - & - & - & - \\ 
			25 & - & - & - & - & - & 0.065 & - & - & - & - & - & 0.478 \\ 
			\hline
		\end{tabular} 
	}
	\caption{Annualized Sharpe ratio and standard error of the Sharpe ratio for every model and each group number $g$. Standard errors are calculated using standard bootstrap samples with 1000 repetitions. ``-'' indicates that there was no model calculated for that number of groups.}
	\label{tab:results}
\end{table}

The results of our empirical study are shown both in Figure~\ref{fig:res} as well as Table~\ref{tab:results}. The orange dots in the figure represent the annualized Sharpe ratio of our proposed model $TS_g$ for all numbers of groups we have specified, for example, $g=5$ means that the dataset was split into exactly five groups. The black curve between the orange dots is a trend curve, fitted with b-splines, which shows the general development of the performance for each number of groups.\footnote{The trend curve can be fitted in many ways and hence serves only an illustrative purpose.} The gray dots represent the corresponding model but with randomly classified companies. The bars beneath the dots show the difference of our approach compared to the random benchmark $RND_g$. The respective numerical values can be obtained from the right $y$-axis. If the bar is filled with blue color, the proposed model performs better, if it is red, the random model exhibits a higher Sharpe ratio. The dashed lines represent the Sharpe ratio of our other four benchmarks: $MW_{full}$ (red), $N_{full}$ (salmon), $TR2$ (blue) and $TR4$ (green).

By analyzing Figure~\ref{fig:res}, we can observe that our approach of t-SNE in combination with spectral clustering $TS_g$ has an interesting development of the Sharpe ratio, depending on the number of groups. First, the Sharpe ratio seems to be quite low. However, increasing the number of groups leads to a steady increase in the performance measure up to a certain number of groups and forms a maximum. Starting with six groups, the Sharpe ratio declines and moves towards the performance of the tangency portfolio. The gray points of the random portfolio seem to slowly move towards the Sharpe ratio of the tangency portfolio as well when the number of groups increases. This behavior, however, is to be expected, as increasing the number of groups drives the results more and more towards the standard scenario, where each company represents a group. In this case, using the before-mentioned scheme leads to the standard tangency portfolio, as the naive portfolio of one company is the company itself. In accordance with standard portfolio theory (the more assets are included, the better) and the resulting diversification effect, the random grouping strategy $RND_g$ performs overall worse than the tangency portfolio. Nonetheless, $RND_g$ generally outperforms the strategies $TR2$ and $TR4$, which exhibit a constant number of groups per construction (10 and 25, respectively). This result implies that the classification based on Thomson Reuters industry codes does not properly fit the problem at hand.   

Overall, the portfolios based on t-SNE and spectral clustering outperform the tangency portfolio in almost all cases. By applying grouping into naive portfolios the estimation error within $TS_g$ is reduced, which compensates for the lost diversification. On the contrary, the tangency portfolio, although optimally diversified, exhibits high estimation error. While in our set-up, the Markowitz portfolio optimization outperforms the naive portfolio in terms of Sharpe ratio, the outperformance does not seem strong. Therefore, our results support related literature on the importance of the naive portfolio as a suitable benchmark. The portfolios based on industry classification, on the other hand, seem to have no real impact.

Table \ref{tab:results} provides further information on the true annualized Sharpe ratio an investor would have received when utilizing our or one of the benchmark strategies. While the left-hand side shows the Sharpe ratio, the right-hand side displays the standard error associated with that Sharpe ratio, constructed by a standard non-parametric bootstrap with replacement and 1000 repetitions. Analyzing the different standard errors of the portfolios shows a general problem with the statistical inference in the case of Sharpe ratios. Even with thousands of observations the Sharpe ratio still suffers from high standard errors and hence often lacks statistical significance. However, the difference between the best portfolio with our approach $TS_6$ and the tangency portfolio $MW_{full}$ is different on a chosen $\alpha$-level of $0.05$. Aiming at acceptable statistical significance, further comparisons of the multiple Sharpe ratios between the $TS_g$ and $RND_g$ portfolios is not achievable because of the high standard errors and potential stacking of $\alpha$-type errors. 

\section{Discussion}\label{sec5:concl}

In this paper, we introduce a novel approach for classifying financial data. To do so, we utilize a standard machine learning tool, combined with a clustering algorithm. The performed analysis emphasizes the importance of this framework for practitioners with expert knowledge as well as for researchers. While practitioners can use the t-SNE technique to visualize high-dimensional data in a comprehensive way and enrich further analysis with their expert knowledge, researchers can apply t-SNE in combination with the spectral clustering algorithm for a data-driven and efficient company classification.

Furthermore, we create a general decision engine which not only shows the implementation steps for the proposed machine learning approach, but additionally opens up standard approaches such as portfolio optimization for the incorporation of cross-validation and data-driven detection of optimal parameters. Hence, by introducing new parameters for grouping, standard models can be fine-tuned with machine learning. The flexibility of the approach is additionally ensured since the whole model set-up from classification to, for example, optimization can be constructed specifically for the problem at hand. This allows the modeler to create company classifications tailor-made for the problem at hand.

Our article contributes to the existing literature of clustering approaches by combining a dimensionality reduction algorithm with a sophisticated clustering method. By transforming the data into a lower-dimensional space, our approach allows the modeler to tackle the computational problems typically involved when directly applying such clustering algorithms. This is beneficial especially when the implemented techniques require the tuning of parameters and thus a thorough setup of a cross-validation procedure, which is time-consuming on its own. The time saved by our combined approach can be therefore fully allocated to exhaustive grid-based searches for these parameters and thus can lead to more precise results for the underlying area of interest.

The demonstrated effective company classification with machine learning can be used as a substitute for potentially proprietary classification systems and can help identify company similarities according to various measures. Even though we have used return data in our empirical example, the method can be utilized for other data types, such as balance-sheet data or sustainability measures. The empirical example from the field of portfolio optimization shows that t-SNE in combination with spectral clustering significantly outperforms standard benchmarks, such as the tangency or the naive portfolio. Still, our approach is not limited to portfolio theory. We believe that the financial research, as well as the practice can benefit from more applications of our framework to other related topics in finance, such as company valuation or factor analysis.

\bibliographystyle{apalike}
\bibliography{Company_classification_using_machine_learning.bib} 

\clearpage

\end{document}